\documentclass[
  pre,
  twocolumn,
  twoside,
  byrevtex,
  superscriptaddress,
  floatfix,
  nofootinbib,
  longbibliography
]{revtex4-1}

\usepackage[realmainfile]{currfile}
\input{\currfilebase.settings}
\usepackage{url}
\PassOptionsToPackage{hyphens}{url}\usepackage{hyperref}
\makeatletter
\g@addto@macro{\UrlBreaks}{\UrlOrds}
\makeatother

\usepackage{hyperref}
\hypersetup{
  colorlinks=true,
  allcolors=todoblue,
  urlcolor=todoblue,
  citecolor=todoblue,
  pdfborder={0 0 0},
  breaklinks=true,
}

\usepackage{colortbl}

\makeatletter

\def\CT@@do@color{%
  \global\let\CT@do@color\relax
  \@tempdima\wd\z@
  \advance\@tempdima\@tempdimb
  \advance\@tempdima\@tempdimc
  \advance\@tempdimb\tabcolsep
  \advance\@tempdimc\tabcolsep
  \advance\@tempdima2\tabcolsep
  \kern-\@tempdimb
  \leaders\vrule
  \hskip\@tempdima\@plus  1fill
  \kern-\@tempdimc
  \hskip-\wd\z@ \@plus -1fill }
\makeatother

\usepackage{xcolor}

\definecolor{olivegreen}{rgb}{0.33333,.41961,0.18431}
\definecolor{forestgreen}{rgb}{0.13333,.5451,0.13333}
\definecolor{lightgrey}{rgb}{0.7,0.7,0.7}
\definecolor{verylightgrey}{rgb}{0.90,0.90,0.90}
\definecolor{veryverylightgrey}{rgb}{0.95,0.95,0.95}
\definecolor{grey}{rgb}{0.5,0.5,0.5}

\definecolor{headerblue}{HTML}{33367E}
\definecolor{unitednationsblue}{HTML}{4D88FF}

\definecolor{charcoal}{HTML}{36454F}
\definecolor{cinerous}{HTML}{98817B}
\definecolor{feldgrau}{HTML}{4D5D53}
\definecolor{glaucous}{HTML}{6082B6}
\definecolor{arsenic}{HTML}{3B444B}
\definecolor{xanadu}{HTML}{738678}

\definecolor{firebrick}{HTML}{B22222}
\definecolor{orangered}{HTML}{FF4500}
\definecolor{tomato}{HTML}{FF6347}

\definecolor{purpletaupe}{HTML}{3B444B}

\definecolor{todoblue}{RGB}{0, 91, 187}

\usepackage{graphicx,epsfig,verbatim,enumerate}
\usepackage{amssymb,amsmath}
\usepackage{ifthen}

\usepackage{tabularx}
\usepackage{multirow, makecell}
\newcolumntype{C}{>{\centering\arraybackslash}X}
\newcolumntype{L}{>{\raggedright\arraybackslash}X}
\newcolumntype{R}{>{\raggedleft\arraybackslash}X}
\usepackage{longtable}

\usepackage{mathtools}

\newboolean{twocolswitch}

\newcommand{\sindex}[1]{}
\newcommand{\nindex}[1]{}

\newcommand{\www}[1]{\url{#1}}

\usepackage{lettrine}

\usepackage{tikz}
\usetikzlibrary{arrows.meta,decorations.pathreplacing}
\usepackage{mathrsfs}
\newtheorem{thm}{Theorem}[section]

\newtheorem{defn}[thm]{Definition}
\newtheorem{rem}[thm]{Remark}
\newtheorem{example}[thm]{Example}

\newcommand{\norm}[1]{\left\lVert #1 \right\rVert}

\setboolean{twocolswitch}{true}
\begin{document}

\title{\protect
  Blending search queries with social media data to improve \\
forecasts of economic indicators


}

\author{
\firstname{Yi}
\surname{Li}
}
\email{yli63@massmutual.com}

\author{
\firstname{Asieh}
\surname{Ahani}
}

\author{
\firstname{Haimao}
\surname{Zhan}
}

\author{
\firstname{Kevin}
\surname{Foley}
}

\affiliation{
  MassMutual Data Science, Amherst, MA 01002
}

\author{
\firstname{Thayer}
\surname{Alshaabi}
}

\affiliation{
  Vermont Complex Systems Center, University of Vermont, Burlington, VT 05401.
}

\affiliation{
  MassMutual Center of Excellence for Complex Systems \& Data Science, Burlington, VT 05401
}

\affiliation{
  Department of Computer Science, University of Vermont, Burlington, VT 05401.
}

\author{
\firstname{Kelsey}
\surname{Linnell}
}

\affiliation{
  Vermont Complex Systems Center, University of Vermont, Burlington, VT 05401.
}

\affiliation{
  MassMutual Center of Excellence for Complex Systems \& Data Science, Burlington, VT 05401
}

\affiliation{
  Department of Mathematics \& Statistics, University of Vermont, Burlington, VT 05401.
}

\author{
\firstname{Peter Sheridan}
\surname{Dodds}
}

\affiliation{
  Vermont Complex Systems Center, University of Vermont, Burlington, VT 05401.
}

\affiliation{
  MassMutual Center of Excellence for Complex Systems \& Data Science, Burlington, VT 05401
}

\affiliation{
  Department of Computer Science, University of Vermont, Burlington, VT 05401.
}

\author{
\firstname{Christopher M.}
\surname{Danforth}
}

\affiliation{
  Vermont Complex Systems Center, University of Vermont, Burlington, VT 05401.
}

\affiliation{
  MassMutual Center of Excellence for Complex Systems \& Data Science, Burlington, VT 05401
}

\affiliation{
  Department of Mathematics \& Statistics, University of Vermont, Burlington, VT 05401.
}

\author{
\firstname{Adam}
\surname{Fox}
}

\affiliation{
  MassMutual Data Science, Amherst, MA 01002
}

\affiliation{
  Vermont Complex Systems Center, University of Vermont, Burlington, VT 05401.
}

\date{\today}

\begin{abstract}
  \protect
  The forecasting of political, economic, and public health indicators using internet activity has demonstrated mixed results.
For example, 
while some measures of explicitly surveyed public opinion 
correlate well with social media proxies, 
the opportunity for profitable investment strategies 
to be driven solely by sentiment extracted from
social media
appears to have expired.
Nevertheless, the internet's 
space of potentially predictive input signals
is combinatorially vast and will continue to 
invite careful exploration.
Here, we combine unemployment related search data from Google Trends with economic language on Twitter to attempt to 
nowcast and forecast: 
1. State and national unemployment claims for the US,
and
2. Consumer confidence in G7 countries.
Building off of a recently developed 
search-query-based model,
we show that
incorporating Twitter data 
improves forecasting of unemployment claims,
while the original method remains marginally better at nowcasting. Enriching the input signal with temporal statistical features (e.g., moving average and rate of change) 
further reduces errors,
and improves the predictive utility of the proposed method when applied to other economic indices, such as consumer confidence.







\end{abstract}



\maketitle


\section{Introduction}\label{sec:pei.introduction}

Macroeconomic time series play a significant role in investment analysis~\cite{indicator2012}, government budget planning~\cite{premchand_public_2001}, and economic forecasting~\cite{tsay_analysis_2010, tsay_multivariate_2014}. Many business stakeholders make decisions based on the projected outputs generated by statistical models of related economic time series. 
As a result, there is a vast literature on time series modeling of economic indicators, ranging in complexity from linear to nonlinear~\cite{tsay_nonlinear_2019}, univariate to multivariate~\cite{tsay_multivariate_2014}, to deep learning~\cite{mlstock}. 

Among all economic measures, leading indicators comprise the subset that usually, but not always, change before the economy as a whole changes~\cite{Ozyildirim2011,osullivan_economics:_2004}. 
For example, weekly unemployment insurance claims data\footnote{\url{https://www.dol.gov/}} are used in current economic analysis of unemployment trends both nationally and in each state~\cite{Ozyildirim2011}. 
Specifically,
\textit{initial claims} measure emerging unemployment and \textit{continued claims} measure the number of persons claiming unemployment benefits for at least a week. 
As another example, the \textit{consumer confidence index} (CCI) provides an indication of future household consumption and savings~\cite{Ozyildirim2011}. 
CCI is based upon survey responses regarding expected financial health, sentiment about the general economic situation, unemployment, and capacity to save. 
The Organization for Economic Co-operation and Development (OECD) regularly provides updates of these and related indices on a monthly basis for multiple countries\footnote{\url{https://data.oecd.org/}}.  

Complementing these traditional measures of economic activity and outlook, alternative internet trace data have gained increasing attention in both the academic literature and the data science industry over the past several years.
The volume, velocity, and variety of data reflecting individual attitudes and opinions available online is hard to fathom.
For example, Google Trends\footnote{\url{https://trends.google.com/trends/}} provides access to an anonymous and aggregated but largely unfiltered sample of actual search requests. 
This allows us to infer public interest in most topics at geographic scales from city to state to country.
Researchers have been utilizing such data in epidemiology~\cite{Yang14473}, biology~\cite{Yang2017} and electronic health records (EHR)~\cite{yang_using_2017}, among others. 
Similarly, social media data has also been extensively investigated in research on ecological economics~\cite{schwartz_visitors_2019}, emotional indicators in slang~\cite{gray_hahahahaha_2019}, and sociotechnical time series~\cite{dewhurst_shocklet_2020, alshaabi2020storywrangler}.

There are several existing statistical and time series models that can be used for time series forecasting. For example, Scott et. al.  \cite{Scott2014,Scott} propose to use Bayesian Structural Time Series (BSTS) to predict economic variables. De Livera et. al. \cite{DeLivera2011} introduce two forecasting methods, BATS and TBATS, which use exponential smoothing and are based on innovations in state-space modeling. 
Recently, Yi et. al. \cite{yi_forecasting_2021} proposed a novel statistical model, the Penalized
Regression with Inferred Seasonality Module (PRISM), to forecast US unemployment claims based on Google Trends data. 
PRISM employs a time series seasonal decomposition and related Google Trends search data as features. 
Li et. al. use LASSO~\cite{hastie2015, bickel2009, bunea2007} to make predictions with lead times spanning from the present (known as \textit{future zero} or \textit{nowcasting}) to three weeks into the future for US unemployment claims data. 

\begin{figure*}[htpb]
  \centering	
    \includegraphics[width=\textwidth]{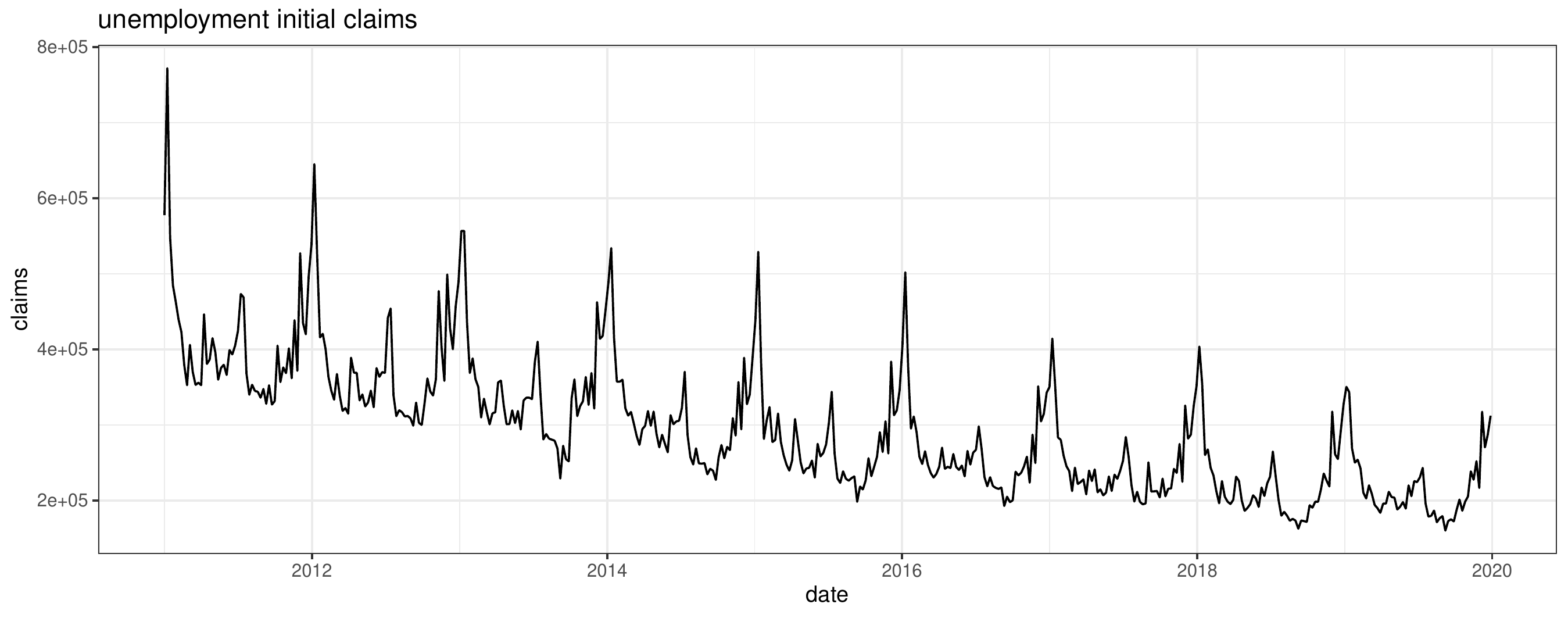}  
  \caption{Weekly unemployment initial claim data (2011 to 2019) collected and aggregated to the national level by the US Department of Labor from each state's unemployment insurance program office. This is a vital and frequently leveraged macroeconomic indicator. However, the data is reported with a one week delay, and thus there is a strong need to predict the current unemployment rate, i.e. nowcast.}
  \label{fig:eda_ic_usa}
\end{figure*}

Motivated by the recent work of Li et. al., in the present study we introduce a generalized version of their model called GPRISM (generalized PRISM).
To be more concrete, our contribution in this paper is threefold: First, we propose a general modeling framework that incorporates flexible time series operators as predictors~\cite{zumbach_operators_2000}.
Second, we leverage the frequency of economic language appearing on Twitter in the modeling task.
Unlike Google Trends data, which is normalized so as to make relative comparisons between separate terms problematic, raw Twitter data allows for proper statistical estimates of phrase popularity.
Inclusion of Twitter data also extends the available training period beyond five years.
Finally, we expand the use case to a more granular level (US States), which could help potential stakeholders with more fine grained unemployment information. 
We also apply our GPRISM model to the aforementioned consumer confidence indicator (CCI). 
In brief, our results suggest that the proposed model outperforms the original in most cases, suggesting that blending social media and search queries can improve the predictive utility of each signal alone.

The paper is structured as follows. Section~\ref{sec:pei.data} describes the data used in this work. Section~\ref{sec:pei.model} demonstrates the details of the proposed GPRISM model, describing how the features are created and how the model is trained. Numerical results on US and state level unemployment claim data, together with the CCI use case are discussed in section \ref{sec:pei.results}, followed with conclusions in section~\ref{sec:pei.concludingremarks}.  

We adopt the following notation throughout the paper. 
Time series (indexed by $t$) are denoted by boldface letters, e.g., $\mathbf{y}_t$. 
The Euclidean norm of a $p$ dimensional vector $\vec{v}\in\mathbb{R}^p$ is denoted by $\norm{\vec{v}}_2$. The $\ell_1$ norm of a vector $\vec{v}$ is denoted by $\norm{\vec{v}}_1$. $\mathbb{R}$ is the set of real numbers. $\mathbf{Z}$ is the set of integers. $\mathscr{L}[\cdot]$ is the time series operator (that maps one time series to another time series). $\mathbb{R}^n_k$ is the $n$ dimensional space $\{0\}^k\times\mathbb{R}^{n-k}$ with the first $k$ dimensions populated by degenerated single value spaces (to represent the transient part of the discrete series). 


\begin{figure*}
  \centering	
    \includegraphics[width=\textwidth]{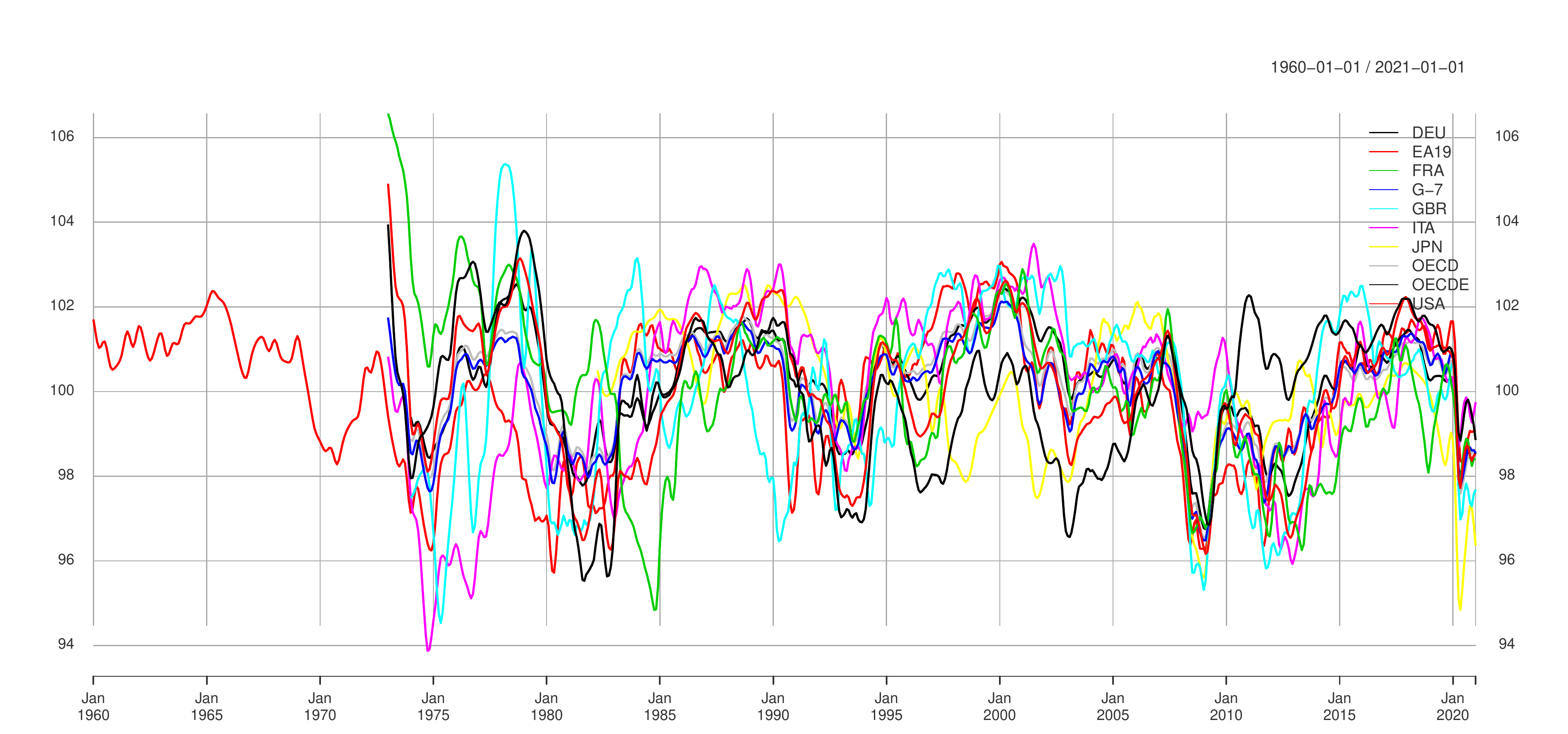}  
  \caption{Monthly Consumer Confidence Index from OECD. An indicator above 100 signals a boost in the consumers’ confidence towards the future economic situation, as a consequence of which they are less prone to save, and more inclined to spend money on major purchases in the next 12 months. Values below 100 indicate a pessimistic attitude towards future developments in the economy, possibly resulting in a tendency to save more and consume less. The lines represent the CCI index for the corresponding country or super-nations Germany (DEU), France (FRA), the United Kingdom (GBR), Italy (ITA), Japan (JPN), the European area (EA19, representing 19 countries); OECD total (OECD), and OECD Europe (OECDE).}
  \label{fig:cci}
\end{figure*}

\section{Data Description}\label{sec:pei.data}

Two main classes of economic time series are considered: US and state level unemployment initial claims data, and country level consumer confidence indicators. Each is described in detail below.

\subsection{Unemployment claims}

An initial unemployment claim is filed by an individual after separation from an employer. 
The claim requests a determination of basic eligibility for the Unemployment Insurance program. 
It is generally known as one of the leading indicators for the economy. The time series data for national unemployment statistics~\cite{yi_forecasting_2021} can be obtained directly from Federal Reserve Economic Data (FRED) \footnote{\url{https://fred.stlouisfed.org/series/ICNSA}}. Figure~\ref{fig:eda_ic_usa} shows the unemployment initial claim data from 2011 to 2019. 
In addition, state level unemployment claim data is available directly from the Employment \& Training Administration (United States Department of Labor) \footnote{\url{https://oui.doleta.gov/unemploy/claims_arch.asp}}, which is the original source of this data set.

\subsection{Consumer confidence index}

The Consumer Confidence Index (CCI) provides a proxy for future household consumption and savings, based upon survey responses regarding expected financial outlook, sentiment about the broader economy, unemployment, and capability of savings. 
Data for the largest economies including G7 countries can be obtained from the OECD website\footnote{\url{data.oecd.org/leadind/consumer-confidence-index-cci.htm}}. Figure~\ref{fig:cci} plots the available CCI data for the 60 year period ending in 2020.

\subsection{Google Trends}

The Google Trends website\footnote{\url{www.google.com/trends}} provides real-time internet search data. The weekly relative popularity of search query terms is available via an API, with the score being normalized to a number between $0$ and $100$. The site can also return other query terms that are most highly correlated with the search term. Here, following Yi et. al., we adopt the top $25$ search terms that are most related to ``unemployment""~ \cite{yi_forecasting_2021}. The detailed terms are provided in Appendix.

\subsection{Twitter}

Social media data reflecting the frequency of specific economic terms is obtained from the Storywrangling project~\cite{alshaabi2020storywrangler}, an open source API offering day scale phrase popularity going back to 2009.\footnote{\url{https://storywrangling.org/}}
Storywrangler collects a random $10\%$ of all public messages using Twitter's Decahose API, and parses tweets into daily frequencies of words, 2-word phrases, and three word expressions, i.e. $n$-grams with $n=1,2,3$.


\section{Model Description}\label{sec:pei.model}


In this section, we give a brief overview of the time series prediction model PRISM~\cite{yi_forecasting_2021}, and then describe the details of our proposed generalization (GPRISM). 
Our main goal is to make predictions of future values of the time series $\{\mathbf{y}_t\}$ using past values of both $\{\mathbf{y}_t\}$, and exogenous time series $\{\mathbf{x}_t\}$.


\begin{figure*}[htpb!]
  \centering
\begin{tikzpicture}[scale=7]
\draw[->, thick] (-1,0) -- (1.1,0);
\foreach \x/\xtext in {-1/$t_{-\infty}$,-0.8/$\cdots$,-0.6/$t_{1-k}$,-0.4/$\cdots$,-0.2/$t_{-1}$,0/$t_0$,0.2/$t_1$,0.4/$t_2$,0.6/$t_3$,0.8/$\cdots$,1/$t_{+\infty}$}
    \draw[thick] (\x,0.5pt) -- (\x,-0.5pt) node[below] {\xtext};
\fill[opacity = 0.5, blue,rounded corners=1ex] (-0.6,-.16ex) -- (0.2, -.16ex) -- (0.2, .16ex) -- (-0.6,.16ex) -- cycle;
\fill[opacity = 0.5, red,rounded corners=1ex] (-0.4,-.32ex) -- (0.4, -.32ex) -- (0.4, .0ex) -- (-0.4,.0ex) -- cycle;
\node at (1.25,0) {$\mathbf{y}_t$};

\draw[->, thick] (-1,-0.2) -- (1.1,-0.2);
\foreach \x/\xtext in {-1/$t_{-\infty}$,-0.8/$\cdots$,-0.6/$t_{1-k}$,-0.4/$\cdots$,-0.2/$t_{-1}$,0/$t_0$,0.2/$t_1$,0.4/$t_2$,0.6/$t_3$,0.8/$\cdots$,1/$t_{+\infty}$}
    \draw[thick] (\x,-5.25pt) -- (\x,-6.25pt) node[below] {\xtext};
\node at (1.25,-0.2) {$\mathbf{\tilde{y}}_t=\mathscr{L}[\mathbf{y}_t]$};

\node at (0.075,-0.125) {\color{blue} $f_{k,\theta}(\cdot)$};
\node at (0.275,-0.125) {\color{red} $f_{k,\theta}(\cdot)$};

\draw[fill=blue, fill opacity=0.5] (0.2,-0.2) circle (0.5pt);
\draw[fill=red, fill opacity=0.5] (0.4,-0.2) circle (0.5pt);
\draw[blue, very thick, ->] (0.1,0)--(0.2,-0.175);
\draw[red, very thick, ->] (0.3,0)--(0.4,-0.175);

\end{tikzpicture}
\caption{Feature creation step with time series operators. For a given time series $\mathbf{y}_t$ and a function $f_{k,\theta}(\cdot)$ with parameters $k$ and $\theta$, we can create a new time series $\tilde{\mathbf{y}}_t$ as a new feature. At each time point in $\mathbf{y}_t$, the previous $k$ elements (red bar) are transformed with $f_{k,\theta}(\cdot)$ into a new value (red dot), which will be the new feature at the corresponding time point. This process is repeated, say for the blue bar and blue dot, to get the whole feature $\tilde{\mathbf{y}}_t$.} 
  \label{fig:step0}
\end{figure*}
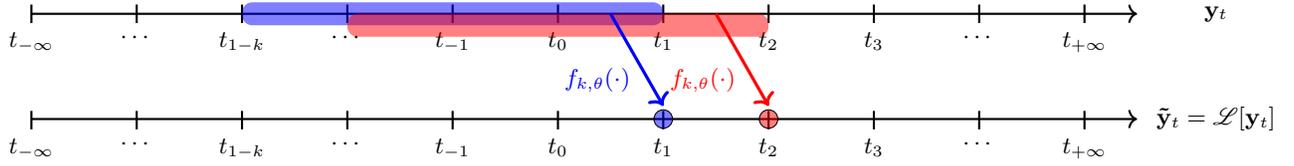

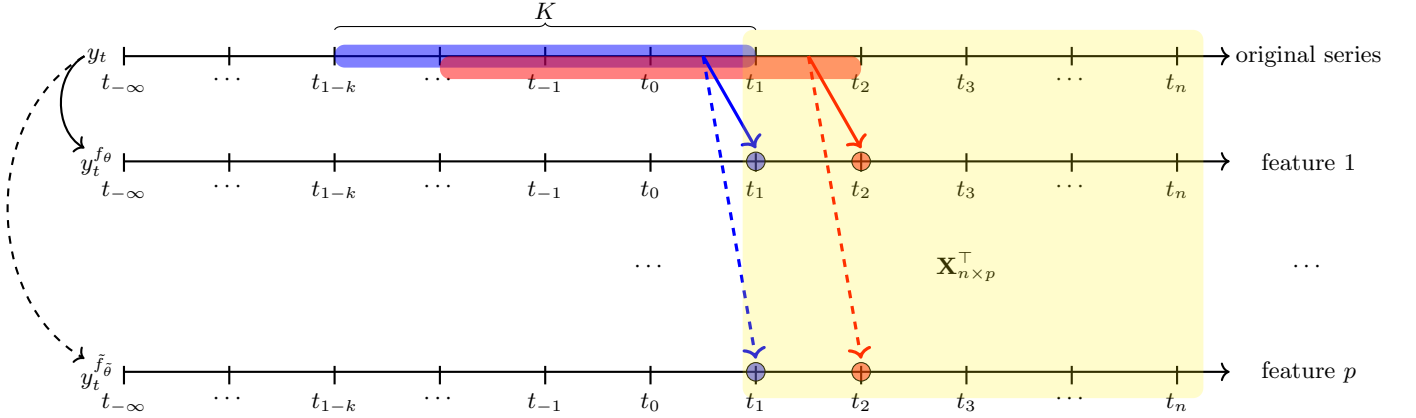
\begin{figure*}[htpb!]
  \centering
\begin{tikzpicture}[scale=7]
\draw[->, thick] (-1,0) -- (1.1,0);
\foreach \x/\xtext in {-1/$t_{-\infty}$,-0.8/$\cdots$,-0.6/$t_{1-k}$,-0.4/$\cdots$,-0.2/$t_{-1}$,0/$t_0$,0.2/$t_1$,0.4/$t_2$,0.6/$t_3$,0.8/$\cdots$,1/$t_n$}
    \draw[thick] (\x,0.5pt) -- (\x,-0.5pt) node[below] {\xtext};
\fill[opacity = 0.5, blue,rounded corners=1ex] (-0.6,-.16ex) -- (0.2, -.16ex) -- (0.2, .16ex) -- (-0.6,.16ex) -- cycle;
\fill[opacity = 0.5, red,rounded corners=1ex] (-0.4,-.32ex) -- (0.4, -.32ex) -- (0.4, .0ex) -- (-0.4,.0ex) -- cycle;
\draw (-1.05,0) node {$y_t$};
\node at (1.25,0) {original series};

\draw[->, thick] (-1,-0.2) -- (1.1,-0.2);
\foreach \x/\xtext in {-1/$t_{-\infty}$,-0.8/$\cdots$,-0.6/$t_{1-k}$,-0.4/$\cdots$,-0.2/$t_{-1}$,0/$t_0$,0.2/$t_1$,0.4/$t_2$,0.6/$t_3$,0.8/$\cdots$,1/$t_n$}
    \draw[thick] (\x,-5.25pt) -- (\x,-6.25pt) node[below] {\xtext};
\draw (-1.05,-0.2) node {$y_t^{f_{\theta}}$};
\node at (1.25,-0.2) {feature $1$};

\draw[decorate,decoration=brace] (-0.6,0.05) -- (0.2,0.05) node[midway,above=0.1em]{$K$};

\node at (0,-0.4) {$\cdots$};
\node at (1.25,-0.4) {$\cdots$};
\node at (0.6,-0.4) {$\mathbf{X}^\top_{n\times p}$};

\draw[->, thick] (-1,-0.6) -- (1.1,-0.6);
\foreach \x/\xtext in {-1/$t_{-\infty}$,-0.8/$\cdots$,-0.6/$t_{1-k}$,-0.4/$\cdots$,-0.2/$t_{-1}$,0/$t_0$,0.2/$t_1$,0.4/$t_2$,0.6/$t_3$,0.8/$\cdots$,1/$t_n$}
    \draw[thick] (\x,-16.75pt) -- (\x,-17.75pt) node[below] {\xtext};
\draw (-1.05,-0.6) node {$y_t^{\tilde{f}_{\tilde{\theta}}}$};
\node at (1.25,-0.6) {feature $p$};

\draw[fill=blue, fill opacity=0.5] (0.2,-0.2) circle (0.5pt);
\draw[fill=red, fill opacity=0.5] (0.4,-0.2) circle (0.5pt);
\draw[blue, very thick, ->] (0.1,0)--(0.2,-0.175);
\draw[red, very thick, ->] (0.3,0)--(0.4,-0.175);
\draw[thick, ->] (-1.075,0) to [out =210, in =150] (-1.075,-0.175);

\draw[fill=blue, fill opacity=0.5] (0.2,-0.6) circle (0.5pt);
\draw[fill=red, fill opacity=0.5] (0.4,-0.6) circle (0.5pt);
\draw[blue, very thick, dashed, ->] (0.1,0)--(0.2,-0.575);
\draw[red, very thick, dashed, ->] (0.3,0)--(0.4,-0.575);
\draw[thick, dashed, ->] (-1.075,0) to [out =210, in =150] (-1.075,-0.575);

\fill[opacity = 0.2, yellow,rounded corners=1ex] (0.175,-0.65) -- (1.05, -0.65) -- (1.05, 0.05) -- (0.175,0.05) -- cycle;

\end{tikzpicture}
\caption{Step 1: feature creation with time series operators. For a given function $f_{\theta}$, we can create a new feature (red, feature 1) with the aforementioned process by applying $f_{\theta}$ on the original time series. Similarly, with additional types of functions (say $\tilde{f}_{\tilde{\theta}}$), we can repeatedly create multiple features (until the $p^{th}$ feature). The whole data matrix can be calculated in this fashion with enough historical data. In the end, we can choose the desired time period as our final modeling data matrix $\mathbf{X}_{n\times p}$.}
  \label{fig:step1}
\end{figure*}

Depending on the application, $\{\mathbf{y}_t\}$ may experience delay or administrative lag due to the time required to process paperwork.
For example, unemployment claim data $\{\mathbf{y}_t\}$ is usually unavailable at time $t$, but estimates are released publicly weeks later.  
Based on the information $\{\mathbf{y}_i\}_{i=1}^{t-1}$, we may also want to perform a nowcast or real-time prediction in addition to the future forecasts, e.g., two weeks ahead. 
However, this does not affect the general mathematical formulation of the time series model\footnote{Note that  we can always shift the time series to meet the application need, as long as future information is not used to train the model.}.  

The first step of the PRISM model is to apply a seasonal decomposition on the previous fixed window of length $M$. 
Specifically, at time $t$ we decompose the time series $\{\mathbf{y}_i\}_{i=t-M}^{t-1}$ as follow:

\begin{equation}
\{\mathbf{y}_i\}_{i=t-M}^{t-1}    \xrightarrow[]{\text{seasonal decomposition}} \{\gamma_{i,t}, z_{i,t}\}_{i=t-M}^{t-1}.
\end{equation}

In this expression, $\{\gamma_{i,t}\}_{i=t-M}^{t-1}$ is the estimated seasonal component from the decomposition, while $\{z_{i,t}\}_{i=t-M}^{t-1}$ represents the remaining contributions including trends and noise. 
Note that for a given time $t$, $2M$ additional data points are generated by this process.

The second step is to train the model with penalized regression using the linear model:
\begin{equation}
\mathbf{y}_{t+l} = \mu_y^{(l)} +\sum_{j=1}^K\alpha_j^{(l)} \mathbf{z}_{t-j,t}+ \sum_{j=1}^K\delta_j^{(l)} \mathbf{\gamma}_{t-j,t} + \sum_{i=1}^q\beta_i^{(l)} \mathbf{x}_{i,t}. 
\end{equation}

Here, $K$ is how many lagged historical decomposed data points we want to use, and $\alpha, \delta, \beta$ are fitted coefficients for the non-seasonal, seasonal and exogenous features. 
Note that for each prediction length $l$, we train a different model.
As a result, coefficients are labeled with $(l)$ to indicate separate model predictions at time $t$. 
The penalization terms with $l_1$ penalty for the coefficients are added to the objective function, as for the normal LASSO model\footnote{which can be trained with the $\mathtt{glmnet}$ package in $\mathbf{R}$.}. We also note that the time resolution (for a single time step) in our applications are weekly for the unemployment claims and monthly for the consumer confidence index. 

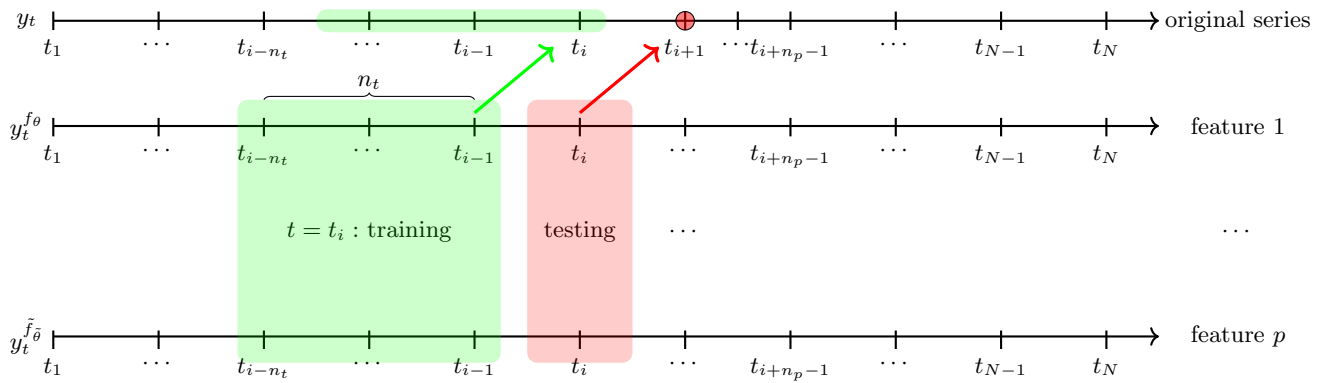
\begin{figure*}[htpb!]
  \centering
\begin{tikzpicture}[scale=7]
\draw[->, thick] (-1,0) -- (1.1,0);
\foreach \x/\xtext in {-1/$t_1$,-0.8/$\cdots$,-0.6/$t_{i-n_t}$,-0.4/$\cdots$,-0.2/$t_{i-1}$,0/$t_{i}$,0.2/$t_{i+1}$,0.3/$\cdots$,0.4/$t_{i+n_p-1}$,0.6/$\cdots$,0.8/$t_{N-1}$,1/$t_N$}
    \draw[thick] (\x,0.5pt) -- (\x,-0.5pt) node[below] {\xtext};
\draw (-1.05,0) node {$y_t$};
\node at (1.25,0) {original series};

\draw[->, thick] (-1,-0.2) -- (1.1,-0.2);
\foreach \x/\xtext in {-1/$t_1$,-0.8/$\cdots$,-0.6/$t_{i-n_t}$,-0.4/$\cdots$,-0.2/$t_{i-1}$,0/$t_{i}$,0.2/$\cdots$,0.4/$t_{i+n_p-1}$,0.6/$\cdots$,0.8/$t_{N-1}$,1/$t_N$}
    \draw[thick] (\x,-5.25pt) -- (\x,-6.25pt) node[below] {\xtext};
\draw (-1.05,-0.2) node {$y_t^{f_{\theta}}$};
\node at (1.25,-0.2) {feature $1$};

\node at (0.2,-0.4) {$\cdots$};
\node at (1.25,-0.4) {$\cdots$};
\node at (-0.4,-0.4) {$t=t_i:$ training};
\node at (0,-0.4) {testing};

\draw[->, thick] (-1,-0.6) -- (1.1,-0.6);
\foreach \x/\xtext in {-1/$t_1$,-0.8/$\cdots$,-0.6/$t_{i-n_t}$,-0.4/$\cdots$,-0.2/$t_{i-1}$,0/$t_{i}$,0.2/$\cdots$,0.4/$t_{i+n_p-1}$,0.6/$\cdots$,0.8/$t_{N-1}$,1/$t_N$}
    \draw[thick] (\x,-16.75pt) -- (\x,-17.75pt) node[below] {\xtext};
\draw (-1.05,-0.6) node {$y_t^{\tilde{f}_{\tilde{\theta}}}$};
\node at (1.25,-0.6) {feature $p$};

\fill[opacity = 0.2, green,rounded corners=1ex] (-0.65,-0.65) -- (-0.15, -0.65) -- (-0.15, -0.15) -- (-0.65,-0.15) -- cycle;
\fill[opacity = 0.2, green,rounded corners=1ex] (-0.5,-.16ex) -- (0.05, -.16ex) -- (0.05, .16ex) -- (-0.5,.16ex) -- cycle;

\fill[opacity = 0.2, red,rounded corners=1ex] (-0.1,-0.65) -- (0.1, -0.65) -- (0.1, -0.15) -- (-0.1,-0.15) -- cycle;
\draw[fill=red, fill opacity=0.5] (0.2,0) circle (0.5pt);
\draw[green, very thick, ->] (-0.2,-0.175) -- (-0.05,-0.05);
\draw[red, very thick, ->] (0,-0.175) -- (0.15,-0.05);

\draw[decorate,decoration=brace] (-0.6,-0.15) -- (-0.2,-0.15) node[midway,above=0.1em]{$n_t$};

\end{tikzpicture}
\caption{Step 2: model training for the one step ahead prediction task. With the modeling data matrix $\mathbf{X}_{n\times p}$, we can create our training/testing and predictor/features pairs on a rolling/walk forward basis. At a given time point, the original time series is shifted ahead with desired steps depending on the need, and the most recent $n_t$ data until the last time point with the $p$ features are selected as the training data, while features from the current time point are used to make the prediction as test set. The model is always updated as new data arrives.}
  \label{fig:step2}
\end{figure*}


\subsection{GPRISM}

To generalized PRISM so that it can incorporate additional temporal information and predictive signals, we start by defining a time series operator.

\begin{defn} [infinite version] Assume $\mathbf{y}_t\in\mathbb{R}^{\mathbf{Z}}$ 
is an infinite time series and define a function $f_{k,\theta}(\cdot):\mathbb{R}^k\to\mathbb{R}$, 
where $k$ is the window size and $\theta$ is the parameter set. 
$f_{k,\theta}(\cdot)$ 
induces a \textbf{time series operator} $\mathscr{L}^{f_{k,\theta}}[\cdot]:\mathbb{R}^{\mathbf{Z}}\to\mathbb{R}^{\mathbf{Z}}$ 
that maps $\mathbf{y}_t$ to $\mathbf{\tilde{y}}_t\in\mathbb{R}^{\mathbf{Z}}$ s.t.
\begin{equation}
    \mathbf{\tilde{y}}_i = f_{k,\theta}(\mathbf{y}_{(i-k+1):i})
\end{equation}
i.e. $\mathbf{\tilde{y}}_t=\mathscr{L}^{f_{k,\theta}}[\mathbf{y}]_t$.
\end{defn}

In practice, time series observations are finite. 
As long as we have enough historical data, we can ignore the first transient part of the discrete series\footnote{ This can be as short as length $k-1$, where $k$ is the window size}.   

\begin{figure*}[htpb]
  \centering	
    \includegraphics[width=\textwidth]{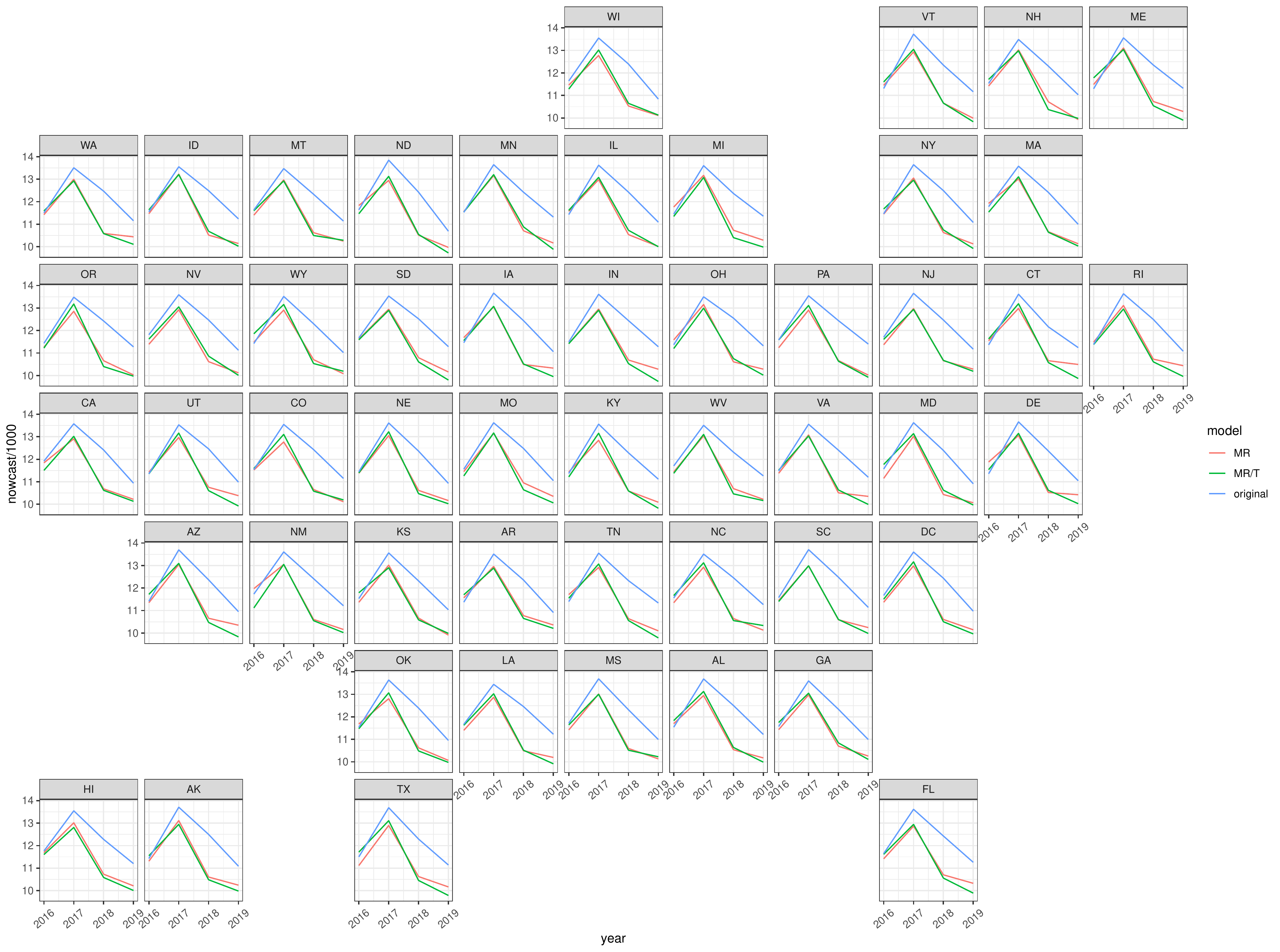}  
  \caption{The state level unemployment initial claim results for the nowcast task. The x-axis is the years and the y-axis is the prediction MAE results in thousands. Three models are compared here: the original model (blue), the model with additional moving average and rate of change features (red), and the model with additional features together with the Twitter data. }
  \label{fig:mae_nowcast}
\end{figure*}

\begin{defn} [finite version] Assume $\mathbf{y}_t\in\mathbb{R}^{n}$ 
is a finite time series with length $n$ and define a function $f_{k,\theta}(\cdot):\mathbb{R}^k\to\mathbb{R}$, 
where $k$ is the window size and $\theta$ is the parameter set. $f_{k,\theta}(\cdot)$
induces a \textbf{time series operator} $\mathscr{L}^{f_{k,\theta}}[\cdot]:\mathbb{R}^{n}\to\mathbb{R}_{k-1}^{n}$ that maps $\mathbf{y}_t$ to $\mathbf{\tilde{y}}_t\in\mathbb{R}_{k-1}^{n}$
s.t.
\begin{equation}
    \mathbf{\tilde{y}}_i = f_{k,\theta}(\mathbf{y}_{(i-k+1):i})
\end{equation}
for $i\ge k$ and $\mathbf{\tilde{y}}_i=0$ (or other dummy placeholder) for $i<k$, i.e. $\mathbf{\tilde{y}}_t=\mathscr{L}^{f_{k,\theta}}[\mathbf{y}]_t$.
\end{defn}

Note that the window size $k$ can also be viewed as a parameter for the function $f(\cdot)$. 
Since it is a common parameter for most of the operators, we separate it from $\theta$, which can vary depending on the operator type. 
One can create specific operators depending on the nature of the task, below are some examples of those we commonly use.

\begin{example}[smoothing] Let $\mathbf{x}\in\mathbb{R}^k$. Arguably the easiest information one can extract from a time series would be the moving average, e.g. 
\begin{equation}
    f_{k}(\mathbf{x}) = \frac{1}{k}\sum_{i=1}^kx_i.
\end{equation}
\end{example}

\begin{rem} Based on the nature of the data, one should be careful about whether the last several days of information are appropriate as features. 
For example, in the unemployment claim application, real-time $t$ information is not available. 
Thus, when crafting the function $f$, one should be aware of this issue to avoid look ahead bias, i.e.

\begin{equation}
    f_{k}(\mathbf{x}) = \frac{1}{k-1}\sum_{i=1}^{k-1}x_i.
\end{equation}
and similar for other operators.  
\end{rem}

\begin{example}[stabilization] Let $\mathbf{x}\in\mathbb{R}^k$. One way to extract time series information from historical data is by measuring its rate of change from baseline. 
This operation can be generated with the following function:
\begin{equation}
    f_{k}(\mathbf{x}) = \frac{x_k-x_1}{x_1}.
\end{equation}
\end{example}

\begin{example}[seasonal decomposition] 
Let $\mathbf{x}\in\mathbb{R}^k$ and $\mathsf{SD}$ represent a seasonal decomposition. 
This process results in $\gamma$ (and $z$) of the input time series with the original length, e.g., $k$. 
Let $f_{k,j}$ be the function that only takes the $j^{th}$ element of the decomposition results: 
\begin{equation}
    f_{k,j}(\mathbf{x}) = \mathsf{SD}(\mathbf{x})_j.
\end{equation}
Thus, a series of functions $\{f_{k,j}\}_{j=1}^{M}$ defines the desired features in PRISM.  
\end{example}

Furthermore, indicators from traditional technical analysis in finance~\cite{Lo2000} can also be viewed as examples of time series operators. 
We now move on to a discussion about how to utilize these operators to engineer features and train the model. 

\subsection{Feature creation}

For simplicity, let us first focus on the case in which we only have one function/operator $f/\mathscr{L}$ for a series $\mathbf{y}_t$. 
Assuming we have enough historical data, Figure~\ref{fig:step0} suggests that by applying $f$ repeatedly on a rolling basis over time, a second time series $\mathbf{\tilde{y}}_t$ can be generated with $\mathbf{\tilde{y}}_t=\mathscr{L}[\mathbf{y}_t]$ as a new feature that incorporates previous temporal information defined with $f$.

For example, at time $t=t_1$, our goal is to create a new feature for $t=t_1$ (blue circle in the second axis) using the original timeseries $\mathbf{y}_t$. 
We learn the function $f$ on the original time series using a window length of $k$ (blue strip in the first axis) using the desired value (blue circle) as a label.
For the next time point, we repeat this process for the red strip and red circle, and so on.

This process can be applied with arbitrary type/number ($p$) of time series operators. 
Then, as is shown in Figure~\ref{fig:step1}, $p$ new features can be generated based on the original time series. 
For example, the solid blue and red arrows represent the process we described above for one function $f$. 
With a separate function $\tilde{f}$ (dashed arrows), another feature can be created in the same fashion. 
By trimming the initial transient period data, we arrive at the final data matrix $\mathbf{X}_{n\times p}^\top$, with sample size $n$ and dimension $p$.

\subsection{Model training}

With the data matrix in place (including the corresponding exogenous data if needed), we can proceed to train the model on a rolling/walk forward basis. 
Figure~\ref{fig:step2} illustrates how to use historical data up to time point $t$ to train the model, and make $l$-step predictions (in the figure, $l=1$). 
For example, at time $t=i$, the task is to predict the next original time point $t=i+1$ using the features at $t=i$. 
So this pair of data becomes our testing set. 
On the other hand, we use all the data available up to time $t=i-1$ (for the features) to train the model.

\subsection{Evaluation Metrics}

Prediction results are evaluated in the test region using the root mean squared error (RMSE)
\begin{equation}
    RMSE(\mathbf{\hat{y}}_t,\mathbf{y}_t) = \frac{1}{\sqrt{n_{\text{test}}}} \norm{\mathbf{\hat{y}}_t-\mathbf{y}_t}_2,
\end{equation}
and mean absolute error (MAE)
\begin{equation}
    MAE(\mathbf{\hat{y}}_t,\mathbf{y}_t) = \frac{1}{n_{\text{test}}} \norm{\mathbf{\hat{y}}_t-\mathbf{y}_t}_1,
\end{equation}

\section{Results}\label{sec:pei.results}

Having described the data and model, we now provide the numerical results associated with the model performance on the aforementioned data sets.

\subsection{Unemployment claim and social media data}

In the unemployment claim task, we first present the performance of our GPRISM model on national US data before refining the experiments to the individual state level.
We download statistics on relative popularity for the key words suggested by prior work on Google Trends~\cite{yi_forecasting_2021} and further filter out $n$-grams with a high missing data rate. 
(See the Appendix for a reference table of terms).
For Twitter data, we use frequencies associated with the terms \texttt{unemployed}, \texttt{unemployment}, and \texttt{unemployment rate}.


The four time ranges considered in this experiment are 2010-2014, 2012-2016, 2014-2018 and 2016-2020.
These ranges are necessitated by one notable drawback of Google Trends data, namely that it is transformed using a different normalization factor in every five year range~\cite{yi_forecasting_2021}. 
We compare the performance of four models at prediction tasks associated with these data sets. 
First, we evaluate the PRISM model without Google Trends data (original). 
Second, we deploy the GPRISM model, where the moving average and rate of change features are included but without Twitter data (MR).
Third, we implement the PRISM model with Google Trends data included (original/G).
Finally, we test the GPRISM model which uses the moving average and rate of change features as well as Twitter data (MR/T). 

Table~\ref{tab:rmse} shows the RMSE results for four different prediction tasks, i.e., $i$ weeks into the future for $i=0,1,2,3$, where $i=0$ corresponds to a nowcast.
Performance for the models without social media data are shown in the first two columns (lowest RMSE marked with asterisk), while the models with the social media data appear in the last two columns (having the lowest RMSE marked with asterisk).
The overall lowest RMSE models are bold faced. 
As we can see from the results, the proposed GPRISM model outperforms the original model in most cases except for the nowcast, 
which suggests that the Google Trend data may be more suitable for instantaneous estimates, while the Twitter data may provide more insights on a longer time horizon.
One potential mechanism at play is that individuals searching for information about unemployment, or searching for a new job, do so privately on Google, while Twitter is the platform to publicly express fear about losing a job in the near future.
Similar conclusion can be drawn from the MAE result, which is provided in Appendix. 

\begin{table}[htpb!]
\footnotesize
\begin{tabularx}{\columnwidth}{|L|C|CC|CC|}
\hline
\rowcolor[HTML]{C0C0C0} 
task                                             & dates & \multicolumn{1}{c}{\cellcolor[HTML]{C0C0C0}original} & MR                   & \multicolumn{1}{c}{\cellcolor[HTML]{C0C0C0}original/G} & MR/T                 \\ \hline
                                                 & 2010-2014  & 28058                                                 & {*25262}          & {*\textbf{21046}}                                    & 25252                \\
                                                 & 2012-2016  & {*\textbf{16318}}                                  & 18156                & 18359                                                   & {*18202}          \\
                                                 & 2014-2018  & 15985                                                 & {*15343}          & {*\textbf{14127}}                                    & 15410                \\
\multirow{-4}{*}{nowcast}                        & 2016-2020  & 15083                                                 & {*14013}          & {*\textbf{13821}}                                    & 14151                \\ \hline
\rowcolor[HTML]{EFEFEF} 
\cellcolor[HTML]{EFEFEF}                         & 2010-2014  & 28756                                                 & {*25697}          & 25932                                                   & {*\textbf{25455}} \\
\rowcolor[HTML]{EFEFEF} 
\cellcolor[HTML]{EFEFEF}                         & 2012-2016  & {*\textbf{16740}}                                  & 18101                & 19439                                                   & {*18085}          \\
\rowcolor[HTML]{EFEFEF} 
\cellcolor[HTML]{EFEFEF}                         & 2014-2018  & 16014                                                 & {*15215}          & 15755                                                   & {*\textbf{15141}} \\
\rowcolor[HTML]{EFEFEF} 
\multirow{-4}{*}{\cellcolor[HTML]{EFEFEF}pred.1} & 2016-2020  & 14725                                                 & {*13905}          & 14264                                                   & {*\textbf{13891}} \\ \hline
                                                 & 2010-2014  & 29194                                                 & {*26479}          & 26944                                                   & {*\textbf{26470}} \\
                                                 & 2012-2016  & {*\textbf{16386}}                                  & 16715                & 18733                                                   & {*16680}          \\
                                                 & 2014-2018  & 15852                                                 & {*15213}          & 15452                                                   & {*\textbf{14987}} \\
\multirow{-4}{*}{pred.2}                         & 2016-2020  & 15125                                                 & {*\textbf{13909}} & 14487                                                   & {*14132}          \\ \hline
\rowcolor[HTML]{EFEFEF} 
\cellcolor[HTML]{EFEFEF}                         & 2010-2014  & 29900                                                 & {*\textbf{26934}} & 27159                                                   & {*27109}          \\
\rowcolor[HTML]{EFEFEF} 
\cellcolor[HTML]{EFEFEF}                         & 2012-2016  & {*\textbf{15897}}                                  & 16706                & 18057                                                   & {*16379}          \\
\rowcolor[HTML]{EFEFEF} 
\cellcolor[HTML]{EFEFEF}                         & 2014-2018  & 16364                                                 & {*\textbf{15409}} & 15814                                                   & {*15700}          \\
\rowcolor[HTML]{EFEFEF} 
\multirow{-4}{*}{\cellcolor[HTML]{EFEFEF}pred.3} & 2016-2020  & 15246                                                 & {*14178}          & 14918                                                   & {*\textbf{13705}} \\ \hline
\end{tabularx}

\caption{The RMSE for the US level unemployment claim results. Four tasks are considered here, i.e. nowcast, future one, two, three weeks ahead prediction with four time ranges (2010-2014, 2012-2016, 2014-2018, and 2016-2020). Moreover, four models are compared here in the last columns, the original model (original), the model with additional moving average and rate of change features (MR), the original model with the Google Trends data (original/G) and the MR model with the Twitter data (MR/T). The first two models do not use the exogenous features, while the last two models use social media features. For each of the simulation scenario (row), the smallest error among the four models is in bold face, while the smaller one among the two models (for either without social media models, and with social media models) is denoted with an asterisk.}
\label{tab:rmse}
\end{table}


In addition to the US national level, we also train and test the model for results at the level of individual states.
We expect that this refined spatial scale will be of great interest for economic and business practitioners to make more accurate decisions.
Figure~\ref{fig:mae_nowcast} presents the MAE of the nowcast model applied on multiple states from 2016 to 2019. 
The PRISM model is shown in blue (original), while the GPRISM model with the moving average and rate of change features is shown in red (MR), while the GPRISM model with additional Twitter data is shown in green (MR/T). 

All models perform worse on the nowcast task in all states for 2017, and best in 2019. The three models exhibit similar performance for 2016 and 2017, but GPRISM improves upon PRISM performance for 2018 and 2019.
For lead times 1-3 weeks into the future, model performances degrade as expected, but the prediction tasks have respectively similar results when looking across the three models. We provide the 3 week prediction results in the Appendix.


\begin{figure}[htpb]
  \centering	
    \includegraphics[scale=0.4]{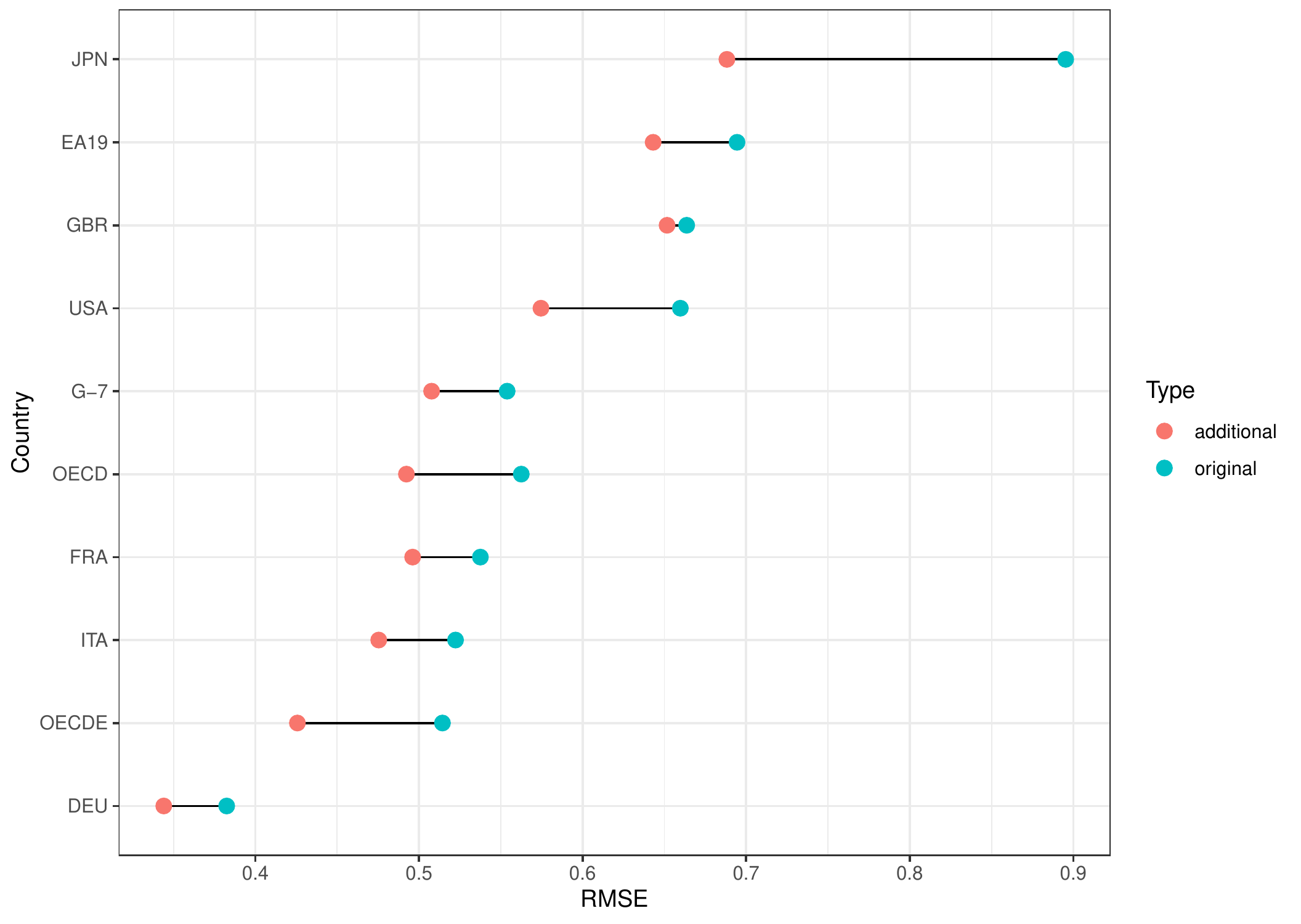}  
  \caption{Errors for the Consumer Confidence Index prediction task across different global indices. Two models are compared here, the original PRISM model (blue) and the model with additional moving average and rate of change features GPRISM (red). The x-axis is the RMSE with the same unit as in Figure~\ref{fig:cci}, and the y-axis delineates the countries and super-nations.}
  \label{fig:cci-res}
\end{figure}

\subsection{Consumer Confidence Index (CCI)}

We apply our model to another leading economic indicator by forecasting the monthly Consumer Confidence Index from 2010 to 2020.
We consider CCI data for large economies, France, Germany, Italy, Japan, the United Kingdom and the United States (G7 countries except Canada which does not have data directly available on the OECD website), as well as several intergovernmental organizations like EA19 Euro area (19 countries, EU19), G7(G-7), OECD - Europe (OECDE) and OECD - Total (OECD). 

Figure~\ref{fig:cci-res} presents the RMSE results for the one month ahead prediction task. 
The blue dots correspond to the RMSE for the PRISM model, and the red dots correspond to the RMSE for the GPRISM model by adding the moving average and rate of change features. 
Our results demonstrate that the proposed GPRISM model yields a lower RMSE.  

\section{Concluding remarks}\label{sec:pei.concludingremarks}

In this paper, we have proposed a general and flexible time series modeling framework with a focus on nowcasting and short term predictions for important economic indicators. 
The framework allows the user to incorporate temporal information derived from historical time series data with various kinds of time series operators, as well as external auxiliary data.
Moreover, the user is able to utilize different machine learning models from the resulting data matrix, which expands possibilities for improving prediction accuracy. 
Social media data is found to enhance the model performance, possibly because relative frequency information is available unlike the Google Trends data.
A pair of important use cases forecasting leading economic indicators demonstrate the preferred model results, which should be of great practical interests for economic decision makers. 

Future directions and applications of this work are vast. 
As one natural extension, model performance out to improve using multivariate time series.
We believe inter correlations between the target variables/series can also provide additional information for the prediction task, and can be exploited as temporal features with time series operators. 
The model can also be further generalized to panel data case, or applied to spatiotemporal applications. 
With additional data from the spatial/instance level dimension, one can increase the resolution of model predictions.

Another natural extension would enable a data-driven selection of words on social media which are found to correlate with economic activity~\cite{dewhurst_shocklet_2020}.
While a small set of economically themed terms were used in the present study, there is potential improvement available if, for example, proxies for fear about losing one's job is found to correlate with future job loss.


\acknowledgments
The authors are grateful for 
support from the Massachusetts Mutual Life Insurance Company,
and for data products made available by Joshua Minot, Michael Arnold, and Jane Adams.

\bibliographystyle{unsrt}
\bibliography{\filenamebase}

\begin{thebibliography}{10}

\bibitem{indicator2012}
Evelina~M. Tainer, editor.
\newblock {\em Using Economic Indicators to Improve Investment Analysis}.
\newblock John Wiley {\&} Sons, Inc., January 2012.

\bibitem{premchand_public_2001}
A.~Premchand.
\newblock Public budgeting and economic development: {E}volution and practice
  of an idea.
\newblock {\em International Journal of Public Administration},
  24(10):1023--1039, 2001.

\bibitem{tsay_analysis_2010}
Ruey~S. Tsay.
\newblock {\em Analysis of financial time series}.
\newblock Wiley series in probability and statistics. Wiley, Cambridge, Mass,
  3rd ed edition, 2010.

\bibitem{tsay_multivariate_2014}
Ruey~S Tsay.
\newblock {\em Multivariate Time Series Analysis: With R and Financial
  Applications}.
\newblock 2014.
\newblock OCLC: 960200458.

\bibitem{tsay_nonlinear_2019}
Ruey~S. Tsay and Rong Chen.
\newblock {\em Nonlinear time series analysis}.
\newblock Wiley series in probability and statistics. John Wiley \& Sons,
  Hoboken, NJ, 2019.

\bibitem{mlstock}
Jingyi Shen and M.~Omair Shafiq.
\newblock Short-term stock market price trend prediction using a comprehensive
  deep learning system.
\newblock {\em Journal of Big Data}, 7(1):66, 2020.

\bibitem{Ozyildirim2011}
Ataman Ozyildirim, Gad Levanon, Brian Schaitkin, and Justyna Zabinska-La
  Monica.
\newblock Comprehensive benchmark revisions for the conference board leading
  economic index{\textregistered} for the united states.
\newblock {\em {SSRN} Electronic Journal}, 2011.

\bibitem{osullivan_economics:_2004}
Economics: principles in action, 2004.
\newblock OCLC: 54052102.

\bibitem{Yang14473}
Shihao Yang, Mauricio Santillana, and S.~C. Kou.
\newblock Accurate estimation of influenza epidemics using google search data
  via argo.
\newblock {\em Proceedings of the National Academy of Sciences},
  112(47):14473--14478, 2015.

\bibitem{Yang2017}
Shihao Yang, Samuel~C. Kou, Fred Lu, John~S. Brownstein, Nicholas Brooke, and
  Mauricio Santillana.
\newblock Advances in using internet searches to track dengue.
\newblock {\em PLOS Computational Biology}, 13(7):1--14, 07 2017.

\bibitem{yang_using_2017}
Shihao Yang, Mauricio Santillana, John~S. Brownstein, Josh Gray, Stewart
  Richardson, and S.~C. Kou.
\newblock Using electronic health records and {Internet} search information for
  accurate influenza forecasting.
\newblock {\em BMC Infectious Diseases}, 17(1):332, December 2017.

\bibitem{schwartz_visitors_2019}
Aaron~J. Schwartz, Peter~Sheridan Dodds, Jarlath P.~M. O'Neil‐Dunne,
  Christopher~M. Danforth, and Taylor~H. Ricketts.
\newblock Visitors to urban greenspace have higher sentiment and lower
  negativity on {Twitter}.
\newblock {\em People and Nature}, 1(4):476--485, December 2019.

\bibitem{gray_hahahahaha_2019}
Tyler~J. Gray, Christopher~M. Danforth, and Peter~Sheridan Dodds.
\newblock Hahahahaha, {Duuuuude}, {Yeeessss}!: {A} two-parameter
  characterization of stretchable words and the dynamics of mistypings and
  misspellings.
\newblock {\em arXiv:1907.03920 [physics]}, July 2019.
\newblock arXiv: 1907.03920.

\bibitem{dewhurst_shocklet_2020}
David~Rushing Dewhurst, Thayer Alshaabi, Dilan Kiley, Michael~V. Arnold,
  Joshua~R. Minot, Christopher~M. Danforth, and Peter~Sheridan Dodds.
\newblock The shocklet transform: a decomposition method for the identification
  of local, mechanism-driven dynamics in sociotechnical time series.
\newblock {\em EPJ Data Science}, 9(1):3, December 2020.

\bibitem{alshaabi2020storywrangler}
Thayer Alshaabi, Jane~L Adams, Michael~V Arnold, Joshua~R Minot, David~R
  Dewhurst, Andrew~J Reagan, Christopher~M Danforth, and Peter~Sheridan Dodds.
\newblock Storywrangler: {A} massive exploratorium for sociolinguistic,
  cultural, socioeconomic, and political timelines using {T}witter, 2021.

\bibitem{Scott2014}
Steven~L. Scott and Hal~R. Varian.
\newblock Predicting the present with bayesian structural time series.
\newblock {\em International Journal of Mathematical Modelling and Numerical
  Optimisation}, 5(1/2):4, 2014.

\bibitem{Scott}
Steven~L. Scott and Hal~R. Varian.
\newblock Chapter 4 - bayesian variable selection for nowcasting economic time
  series / steven l. scott and hal r. varian.
\newblock In {\em Economic Analysis of the Digital Economy}, pages 119--136.
  University of Chicago Press.

\bibitem{DeLivera2011}
Alysha M.~De Livera, Rob~J. Hyndman, and Ralph~D. Snyder.
\newblock Forecasting time series with complex seasonal patterns using
  exponential smoothing.
\newblock {\em Journal of the American Statistical Association},
  106(496):1513--1527, December 2011.

\bibitem{yi_forecasting_2021}
Dingdong Yi, Shaoyang Ning, Chia-Jung Chang, and S.~C. Kou.
\newblock Forecasting unemployment using {Internet} search data via {PRISM}.
\newblock {\em Journal of the American Statistical Association}, 0(ja):1--28,
  2021.

\bibitem{hastie2015}
T.~Hastie, R.~Tibshirani, and M.~Wainwright.
\newblock {\em Statistical Learning with Sparsity: The Lasso and
  Generalizations}.
\newblock Chapman \& Hall/CRC Monographs on Statistics \& Applied Probability.
  Taylor \& Francis, 2015.

\bibitem{bickel2009}
Peter~J. Bickel, Ya’acov Ritov, and Alexandre~B. Tsybakov.
\newblock Simultaneous analysis of lasso and dantzig selector.
\newblock {\em Ann. Statist.}, 37(4):1705--1732, 08 2009.

\bibitem{bunea2007}
Florentina Bunea, Alexandre Tsybakov, and Marten Wegkamp.
\newblock Sparsity oracle inequalities for the lasso.
\newblock {\em Electron. J. Statist.}, 1:169--194, 2007.

\bibitem{zumbach_operators_2000}
Gilles~O. Zumbach and Ulrich~A. Müller.
\newblock Operators on {Inhomogeneous} {Time} {Series}.
\newblock {SSRN} {Scholarly} {Paper} ID 208278, Social Science Research
  Network, Rochester, NY, 2000.

\bibitem{Lo2000}
Andrew~W. Lo, Harry Mamaysky, and Jiang Wang.
\newblock Foundations of technical analysis: Computational algorithms,
  statistical inference, and empirical implementation.
\newblock {\em The Journal of Finance}, 55(4):1705--1765, August 2000.

\end{thebibliography}

\clearpage


\onecolumngrid



\renewcommand{\thefigure}{S\arabic{figure}}
\setcounter{figure}{0}

\renewcommand{\thetable}{S\arabic{table}}
\setcounter{table}{0}

\appendix
\section{Google Trends search terms}

In this part, we provide the terms that is used in the Google Trends data set. Twenty five terms, which are listed in Table~\ref{tab:google}, are used as additional features in the US level unemployment prediction model with Google trends~\cite{yi_forecasting_2021}. 

\begin{table*}[h]
\begin{tabularx}{\textwidth}{CCC}
\hline
unemployment            & unemployment benefits   & unemployment rate    \\
unemployment office     & pa unemployment         & claim unemployment   \\
ny unemployment         & nys unemployment        & ohio unemployment    \\
unemployment florida    & unemployment extension  & texas unemployment   \\
nj unemployment         & unemployment number     & file unemployment    \\
unemployment insurance  & california unemployment & unemployed           \\
unemployment oregon     & new york unemployment   & indiana unemployment \\
unemployment washington & unemployment wisconsin  & unemployment online  \\
unemployment login      &                         &                      \\ \hline
\end{tabularx}
\caption{Top $25$ search query terms related to "unemployment" that is used in the Google Trends data.}
\label{tab:google}
\end{table*}

\subsection{US unemployment MAE}

Table~\ref{tab:mae} shows the results in MAE for four different prediction tasks: nowcast, prediction for future $i$ weeks with $i=1,2,3$. The models without the exogenous social media data are the first two columns (having the lowest MAE marked with asterisk), while the models with the social media data are the last two columns (having the lowest MAE marked with asterisk). The overall lowest MAE are in bold face. As we can see from the results, the proposed model outperforms the origin model in most cases. 

\begin{table}[]
\begin{tabularx}{\columnwidth}{|L|C|CC|CC|}
\hline
\rowcolor[HTML]{C0C0C0} 
task                                             & dates & \multicolumn{1}{c|}{\cellcolor[HTML]{C0C0C0}original} & MR                   & \multicolumn{1}{c|}{\cellcolor[HTML]{C0C0C0}original/G} & MR/T                 \\ \hline
                                                 & 2010-2014  & 23158                                                 & {*21071}          & {*\textbf{16914}}                                    & 21185                \\
                                                 & 2012-2016  & {*\textbf{12140}}                                  & 13158                & 13419                                                   & {*13114}          \\
                                                 & 2014-2018  & 13086                                                 & {*11922}          & {*\textbf{11088}}                                    & 11965                \\
\multirow{-4}{*}{nowcast}                        & 2016-2020  & 11247                                                 & {*10083}          & 10616                                                   & {*\textbf{10076}} \\ \hline
\rowcolor[HTML]{EFEFEF} 
\cellcolor[HTML]{EFEFEF}                         & 2010-2014  & 23758                                                 & {*21463}          & 21634                                                   & {*\textbf{21380}} \\
\rowcolor[HTML]{EFEFEF} 
\cellcolor[HTML]{EFEFEF}                         & 2012-2016  & {*\textbf{12506}}                                  & 12931                & 14482                                                   & {*13025}          \\
\rowcolor[HTML]{EFEFEF} 
\cellcolor[HTML]{EFEFEF}                         & 2014-2018  & 13235                                                 & {*11935}          & 12937                                                   & {*\textbf{11645}} \\
\rowcolor[HTML]{EFEFEF} 
\multirow{-4}{*}{\cellcolor[HTML]{EFEFEF}pred.1} & 2016-2020  & 10862                                                 & {*10327}          & 10332                                                   & {*\textbf{10212}} \\ \hline
                                                 & 2010-2014  & 24021                                                 & {*22144}          & 22642                                                   & {*\textbf{22028}} \\
                                                 & 2012-2016  & {*\textbf{12399}}                                  & 12693                & 14323                                                   & {*12588}          \\
                                                 & 2014-2018  & 13035                                                 & {*11875}          & 12636                                                   & {*\textbf{11769}} \\
\multirow{-4}{*}{pred.2}                         & 2016-2020  & 10965                                                 & {*\textbf{10218}} & 10558                                                   & {*10456}          \\ \hline
\rowcolor[HTML]{EFEFEF} 
\cellcolor[HTML]{EFEFEF}                         & 2010-2014  & 24904                                                 & {*\textbf{22767}} & 23169                                                   & {*22998}          \\
\rowcolor[HTML]{EFEFEF} 
\cellcolor[HTML]{EFEFEF}                         & 2012-2016  & {*\textbf{11941}}                                  & 12690                & 14486                                                   & {*12419}          \\
\rowcolor[HTML]{EFEFEF} 
\cellcolor[HTML]{EFEFEF}                         & 2014-2018  & 13599                                                 & {*\textbf{12320}} & 13288                                                   & {*12654}          \\
\rowcolor[HTML]{EFEFEF} 
\multirow{-4}{*}{\cellcolor[HTML]{EFEFEF}pred.3} & 2016-2020  & 10579                                                 & {*10333}          & 10615                                                   & {*\textbf{9924}}  \\ \hline
\end{tabularx}
\caption{The MAE for the US level unemployment claim results. Four tasks are considered here, i.e. nowcast, future one, two, three weeks ahead prediction with four time ranges (2010-2014, 2012-2016, 2014-2018, and 2016-2020). Moreover, four models are compared here in the last columns, the original model (original), the model with additional moving average and rate of change features (MR), the original model with the Google Trends data (original/G) and the MR model with the Twitter data (MR/T). The first two models do not use the exogenous features, while the last two models use social media features. For each of the simulation scenario (row), the smallest error among the four models is in bold face, while the smaller one among the two models (for either without social media models, and with social media models) is denoted with an asterisk.}
\label{tab:mae}
\end{table}

\section{State level unemployment MAE}

\begin{figure*}[htpb]
  \centering	
    \includegraphics[width=\textwidth]{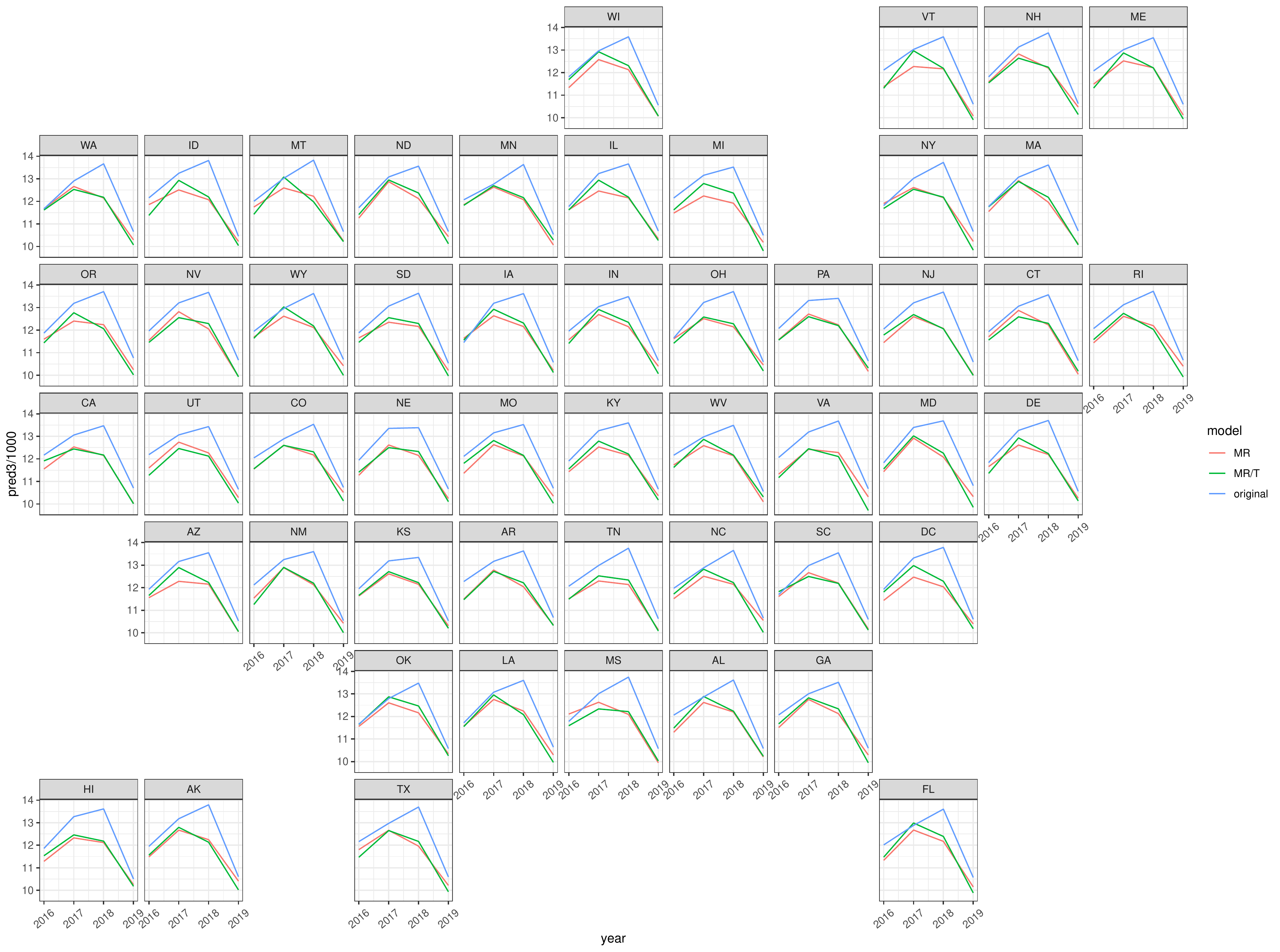}  
  \caption{The state level unemployment initial claim results for the future three-week prediction task. The x-axis is the years and the y-axis is the prediction RMSE results in thousands. Three models are compared here: the original model (blue), the model with additional moving average and rate of change features (red), and the model with additional features together with the Twitter data. }
  \label{fig:mae_pred3}
\end{figure*}

Figure~\ref{fig:mae_pred3} presents the MAE of the models applied on multiples states from 2016 to 2019 for the 3 weeks prediction. The original PRISM model is in red (original), while the GPRISM model by adding moving average and rate of change features is in green (add), while the GPRISM model with additional Twitter data is in blue (add twitter). As we can see from Figure~\ref{fig:mae_nowcast}, the GPRISM model has relatively lower MAE in most of the cases. 

\end{document}